\documentclass[12pt]{article}
\usepackage{amssymb}
\usepackage{amsmath}
\usepackage{graphicx}

\renewcommand{\hbar}{\overline{h}}

\newcommand{\al}{\ensuremath{\alpha}}
\newcommand{\ba}{\begin{align}}
\newcommand{\be}{\begin{equation}}
\newcommand{\bea}{\begin{eqnarray}}

\newcommand{\ea}{\end{align}}
\newcommand{\ee}{\end{equation}}
\newcommand{\eea}{\end{eqnarray}}

\begin{document}

\bigskip
\begin{titlepage}

\begin{flushright}

hep-th/0511106 \\
NORDITA-2005-72
\end{flushright}

\vspace{1cm}

\begin{center}
{\LARGE  The stringy nature of the 2d type-0A black hole \\}

\end{center}
\vspace{3mm}

\begin{center}

{
\bf {Martin  E. Olsson} 
\vspace{5mm}

\it { NORDITA, Blegdamsvej 17\\
DK-2100 Copenhagen, Denmark}

\vspace{5mm}

{\tt
molsson@nordita.dk
}}

\end{center}
\vspace{5mm}

\begin{center}
{\large \bf Abstract}
\end{center}
We investigate the thermodynamics of the RR charged two-dimensional type-0A black hole background at finite temperature, and compare with known 0A matrix model results. It has been claimed that there is a disagreement for the free energy between the spacetime and the dual matrix model. Here we find that this discrepancy is sensitive to how the cutoff is implemented on the spacetime side. In particular, the disagreement is resolved once we put the cutoff at a fixed distance away from the horizon, as opposed to a fixed position in space.
Furthermore, the mass and the entropy of the black hole itself add up to an analytic contribution to the free energy, which is precisely reproduced by the 0A matrix model. We also use results from the 0A matrix model to predict the next to leading order contribution to the entropy of the black hole.
Finally, we note that the black hole is characterized by a Hagedorn growth in its density of states below the Hagedorn temperature. This, together with other results, suggests there is a phase transition at this temperature.

\noindent

\vfill
\begin{flushleft}
November 2005
\end{flushleft}
\end{titlepage}
\newpage

\section{Introduction and motivation}
\bigskip

An important issue in the context of two-dimensional string theories is the question of whether black holes exist in the theory or not. This is important to understand since if they do, they would provide an exciting laboratory for many of the most challenging questions faced in string theory today. For example, what are the microstates making the black hole, how should we think about curvature singularities, and what is the framework for dealing with strongly time-dependent backgrounds such as the interior of the black hole? What makes the 2d type 0 string theories particularly interesting in this regard, is the fact that they have a nonperturbative reformulation in terms of the so called $\hat{c}=1$ matrix models \cite{Douglas:2003up,Takayanagi:2003sm}.

In the case of the 0A string it has been known for some time that the low energy effective theory does indeed predict black hole solutions \cite{Berkovits:2001tg,Gukov:2003yp,Davis:2004xb,Danielsson:2004xf}. These are RR-charged Reissner-Nordstrom like black holes, with a thermodynamics very much like the higher dimensional analogues. Indeed, even the Bekenstein-Hawking entropy-area relation is obeyed by these 2d black holes \cite{Danielsson:2004xf,Davis:2004xi}. Various aspects of these solutions have been studied
\cite{Danielsson:2003yi,Thompson:2003fz,Strominger:2003tm,Ho:2004qp,Aharony:2005hm}.

In the early $c=1$ literature the 0A matrix model, which at the time was known as the deformed matrix model \cite{Jevicki:1993zg}, was conjectured to describe an eternal black hole \cite{Jevicki:1993zg,Danielsson:1993wq}. It is thus very suggestive that the low energy effective theory predicts a charged black hole, with an extremal ground state which indeed should be eternal. A natural thing to do then, as a preliminary check that the matrix model really describes a charged black hole, is to reproduce the black hole thermodynamics directly in the matrix model.

Attempts in this direction have been made. There has been some success, in particular the ground state energy of the entire black hole spacetime is exactly reproduced by the matrix model \cite{Danielsson:2004xf,Davis:2004xi}. However, at nonzero temperature it seems as if the correspondence breaks down \cite{Davis:2004xb,Danielsson:2004xf}.

The origin of the disagreement can be traced to the tree level term of the canonical free energy. It turns out that in the matrix model, the free energy is {\it independent} of temperature \cite{Demeterfi:1993sj,Douglas:2003up,Maldacena:2005he}, while the spacetime analysis suggests it has an explicit temperature dependence \cite{Davis:2004xb,Danielsson:2004xf}. As a result, the matrix model seems to predict a vanishing entropy at tree level, in contradiction with the spacetime result which gives the black hole entropy at the same level. The main purpose of this note is to investigate this discrepancy in more detail.

One important observation is that the temperature dependence on the spacetime side is sensitive to how the volume cutoff is implemented. The volume here is given in terms of the Liouville direction. It is well known that the relation between the Liouville coordinate and the eigenvalue coordinate of the matrix model is subtle, and so far not very well understood. Therefore, one should be careful when expressing quantities on the spacetime side in terms of the coordinates. In particular, in comparing with the matrix model, it is not immediately clear what the most natural way of implementing the cutoff is. Indeed, on the spacetime side one has a freedom of choosing coordinates, while the matrix model on the other hand works in a prefered coordinate system. It is therefore clear that spacetime quantities depending on the choise of coordinate system, can only successfully be matched with matrix model quantities for one specific set of coordinates only.

We will find that in the coordinate system where it is most natural to put the cutoff at a fixed distance away from the horizon, regardless of temperature, the free energy becomes temperature independent. The resulting free energy coincides with the corresponding one on the matrix model side. What is particularly interesting is that, as is clear from the spacetime analysis, the constant free energy is consistent with a non-vanishing entropy.

As a further motivation for these considerations, let us focus for a moment on the free energy at nonzero temperature for the black hole alone. As we will see, the total mass of the spacetime is given by two distinct terms; the energy due to the electrostatic potential as well as the mass of the actual black hole. So, let us express the canonical free energy of the black hole only, i.e.\footnote{We emphasize that this is not the total free energy, which on the spacetime side we denote by $F_{tot}$, where $F_{tot}=F_{BH}+F_{el.stat.}$.}
\be
F_{BH}=M_{BH}-TS_{BH}.
\ee
The black hole, at charge $q$ and temperature $T$, has an entropy given by \cite{Davis:2004xb,Danielsson:2004xf,Davis:2004xi}\footnote{We will state our conventions and derive the thermodynamic quantities in the upcoming section.}
\be
S_{BH}=\frac{q^2}{8}\left(1-\frac{2 \pi T}{k}\right)^{-1}, \qquad k= \sqrt{\frac{2}{\alpha'}},
\ee
and a mass \cite{Davis:2004xi}
\be
M_{BH}=\frac{k}{2\pi}\frac{q^2}{8}\left(1-\frac{2 \pi T}{k}\right)^{-1}.
\ee
Note that the black hole mass is proportional to the entropy, with the proportionality constant given by the limiting, or Hagedorn, temperature $T_{h}=k/2\pi$. This means that the low temperature phase is characterized by a stringy density of states.

We are now ready to write down the free energy for this black hole
\begin{align}\label{eq:bhfree}
F_{BH}&=M_{BH}-TS_{BH}=\frac{k}{2\pi}\frac{q^2}{8}\left(1-\frac{2 \pi T}{k}\right)^{-1}
-T\frac{q^2}{8}\left(1-\frac{2 \pi T}{k}\right)^{-1}\nonumber \\ 
&=(T_{h}-T)\frac{q^2}{8}\left(1-\frac{T}{T_{h}}\right)^{-1}=T_{h}\frac{q^2}{8},
\end{align}
where again $T_{h}=k/2\pi$.
Now we see that the mass and the entropy of the black hole conspire to make the resulting free energy \emph{temperature independent}. Clearly, if the 0A matrix model calculates a temperature independent free energy, it is too soon to conclude that the entropy must vanish. Indeed, the above implies that one should \emph{expect} to get a temperature independent term like (\ref{eq:bhfree}). Interestingly, the term (\ref{eq:bhfree}) is precisely one of the terms the matrix model calculates at tree level. The entire tree level term reads \cite{Danielsson:2004xf}\footnote{The calculation in \cite{Danielsson:2004xf} was done at zero temperature, but, as is well known, the tree level term is unaffected by finite temperature.}
\be \label{eq:mmfree}
F_{MM}=T_{h}\frac{q^2}{8}\left[1-\ln\left(\frac{q^2k^2}{2\epsilon_{c}^2}\right)\right],
\ee
where $\epsilon_{c}$ is an energy cutoff.
Usually the first of the two terms in eq. (\ref{eq:mmfree}) is not included explicitly as it is regarded as an uninteresting analytic term in $q$. But according to what we have just seen, this term is precisely the one which accounts for the black hole. We will see many indications that the black hole hides in the analytic sector of the theory.

The second term in eq. (\ref{eq:mmfree}) is clearly also temperature independent. As mentioned, it has been claimed that this is in contradiction with the spacetime result \cite{Davis:2004xb,Danielsson:2004xf}, which seem to have a temperature dependence there. A purpose of this note is to remedy this discrepancy, and find that the spacetime and matrix model results match perfectly also at finite temperature.

We will only be concerned with the thermodynamics in this paper, and aspects related to it. When we refer to the 0A matrix model we always refer to the singlet sector of the model. Other approaches to find a matrix model black hole in the singlet sector include attempts to create a black hole directly in the matrix model by high energy scattering. This has been done both in the 0B model \cite{Friess:2004tq,Karczmarek:2004bw}, and in the nonextremal limit of the 0A model \cite{Martinec:2004qt}. Also the symmetries of the (bosonic) theory have been carefully explored in this context \cite{Sen:2004yv}.

This paper is organized as follows. We begin in section \ref{sec:geomtherm} by describing the black hole geometry and the corresponding thermodynamics. It is found that the temperature dependence for the free energy is sensitive to how the cutoff is implemented. With a cutoff at a fixed distance away from the horizon, rather than a fixed position in space, the free energy becomes independent of temperature. The coordinate system where this is most naturally realized seem to exhibit some interesting universal features, which are briefly mentioned. We then turn to the matrix model in section \ref{sec:mmbh}, and look at the free energy there. The two constant terms in the spacetime analysis, corresponding to the black hole and the electrostatic potential respectively, are precisely reproduced by the matrix model. From the one-loop term of the matrix model free energy, we predict the next to leading order contribution to the entropy of the black hole. In section (\ref{sec:hagedorn}) we discuss the possibility of a possible phase transition at the Hagedorn temperature. We find three main pieces of evidence for this: the low temperature Hagedorn relation between the mass and the entropy of the black hole, a tachyonic mode appearing precisely at the Hagedorn temperature, and finally a high temperature phase formally similar to the deconfined phase of large $N$ gauge theories. We conclude in section \ref{sec:conclude}.

\bigskip
\section{The black hole geometry and its thermodynamics}\label{sec:geomtherm}
\bigskip

\subsection{The black hole solution}
\bigskip

We begin this section by briefly reviewing known results. This will also serve to set the conventions we will use.

The low energy effective 0A theory we consider here has one RR flux turned on, and is defined by the following action \cite{Berkovits:2001tg}
\be \label{eq:stringact}
 S = - \int \text{d}^2 x \sqrt{-g} \left[ e^{-2\Phi} \left( \mathcal{R} + 4 (\nabla
   \Phi)^2 + \frac{8}{\al '} \right) -\frac{q^2}{4\pi \al '} \right],
\ee
where $\Phi$ is the dilaton field. Furthermore, the tachyon condensate $\langle T(\phi) \rangle =\mu e^{-\phi}$ has been set to zero. That is, we work in a theory where $\mu=0$. At $\mu=0$ it is possible to turn on the second flux $\tilde{q}$. As shown in \cite{Maldacena:2005he} this requires adding $q\tilde{q}$ strings to the spacetime. It was shown there that the result of this in the matrix model, was that the physics in this case only depends on the single parameter $\hat{q}=q+\tilde{q}$ \cite{Maldacena:2005he}. Thus, if we want to generalize our analysis to this case it seems like all we need to do is to dress our $q$ with a hat. This seems to explain the ambiguity of factors of 2 in comparing results with the matrix model \cite{Gukov:2003yp,Davis:2004xb,Danielsson:2004xf,Davis:2004xi}. That is, one can include only one or include both fluxes, the two choices are both compatible with the matrix model.

The equations of motion following from this action are
\begin{align}
\mathcal{R}_{\mu\nu}+\frac{q^2}{8\pi \al '}e^{2\Phi}g_{\mu\nu}+2\nabla_\mu\nabla_\nu\Phi
& =  0, \\
\mathcal{R} + \frac{8}{\al '}+4\nabla^2\Phi-4(\nabla\Phi)^2 & = 0. \label{eq:phieqmot}
\end{align}
Denoting the space-time coordinates by $(t,\phi)$, one finds the
following solution \cite{Berkovits:2001tg,Gukov:2003yp,Davis:2004xb,Danielsson:2004xf,Davis:2004xi}
\begin{align}
 ds^2 & =  - l(\phi) \text{d} t^2 + \frac{1}{l(\phi)} \text{d} \phi^2 \\
\Phi & =  - \sqrt{\frac{2}{\al '}} \phi = -k\phi,
\end{align}
where
\be \label{eq:metriccomp}
 l(\phi) = 1 - \frac{1}{2k} e^{-2k\phi}\left(2k e^{2k \phi_{H}} + \frac{q^2 k^2}{8\pi }(\phi-\phi_{H}) \right),
\ee
where we have been using $k=\sqrt{\frac{2}{\al '}}$. For a given charge $q$, the horizon radius has a minimum $\phi_{ext}$ at extremality. This happens when the two horizons coincide, or, equivalently, when
\be
l(\phi)\Big{|}_{\phi_{ext}}=  l'(\phi)\Big{|}_{\phi=\phi_{ext}}=0.
\ee
This gives
\be \label{eq:phiext}
\phi_{ext}=\frac{1}{2k} \ln \left(\frac{q^2}{32\pi}\right).
\ee
We thus see that for a large value of the flux $q$, the string coupling at the horizon is small
\be
\left.g_{s}\right|_{\phi=\phi_{ext}}=\left.e^{\Phi}\right|_{\phi=\phi_{ext}}=e^{-k\phi_{ext}}=\frac{\sqrt{32\pi}}{q}.
\ee
Furthermore at finite temperature we have $\phi_{H}>\phi_{ext}$, i.e. weaker coupling, so we do not need to worry about quantum corrections to our solution.

We find the temperature of the black hole in the usual way
\be \label{eq:temp}
\frac{1}{\beta}=T=\left. \frac{l'(\phi)}{4\pi}\right|_{\phi_{H}}=\frac{k}{2\pi}\left(1-\frac{q^2}{32\pi}e^{-2k\phi_{H}}\right),
\ee
where $\beta$ denotes the period of euclidean time. Note that the temperature lies in the range $0\leq T\leq k/2\pi$. In particular, we can study the black hole at low temperature. This is unlike the bosonic or uncharged black hole, where the temperature is constant and of order one in string units. This is an important point since non-singlet states of the dual matrix model are expected to be relevant only at high enough temperatures \cite{Gross:1990md} (if at all \cite{Martinec:2004qt}). Therefore, only the singlet sector can be the relevant sector for describing this black hole.

It will become useful later to express the horizon radius in terms of the temperature. Inverting (\ref{eq:temp}) gives
\be
\phi_{H}=\frac{1}{2k}\ln \left(\frac{q^2}{32\pi}\right) - \frac{1}{2k}\ln \left(1-\frac{2\pi T}{k}\right). \label{eq:horiztemp}
\ee
Note that the horizon radius diverges in the limit $T \rightarrow k/2\pi$.

\bigskip
\subsection{The thermodynamics of the black hole}
\bigskip

The ADM mass of this spacetime is given by \cite{Danielsson:2004xf}
\be \label{eq:ADMmass}
M_{tot}=\frac{q^2 k^2}{8\pi }(\phi_{c}-\phi_{H})+ 2k e^{2k \phi_{H}},
\ee
where $\phi_{c}$ is a volume cutoff. Note that this is precisely the coefficient of $e^{-2k\phi}$ in the metric component (\ref{eq:metriccomp}), as it should since it is this term which gives the correction to the asymptotic vacuum. Here we see the first indication that analytic as well as non-analytic terms are playing an important role\footnote{I would like to thank Ulf Danielsson for pointing this out to me.}. The non-analytic dependence on $q$ is implicit in the term linear in $\phi_{H}$, as can be seen from (\ref{eq:phiext}) and (\ref{eq:temp}). Accordingly, the analytic dependence comes from the term exponential in $\phi_{H}$.

The cutoff $\phi_{c}$ in (\ref{eq:ADMmass}) has a clear physical interpretation, and it is important to keep it explicit \cite{Danielsson:2004xf,Davis:2004xi}. Indeed, since our black hole is charged, there is an electrostatic potential and corresponding energy density associated with it. In two spacetime dimensions, the potential is linear in the spatial coordinate so the corresponding energy density must be constant. Thus, the energy becomes proportional to volume. We therefore see that the first of the two terms in (\ref{eq:ADMmass}) is nothing but the electrostatic energy between the volume cutoff and the black hole horizon \cite{Davis:2004xi}. The second term must therefore be the mass of the black hole. Again, note that, after using (\ref{eq:temp}), this term is analytic in $q$. We make this explicit by writing
\begin{align}
M_{el.stat.}&=\frac{q^2 k^2}{8\pi }(\phi_{c}-\phi_{H})=\frac{q^2 k^2}{8\pi }\left(\phi_{c}-\frac{1}{2k}\ln\left[\frac{q^2}{32\pi} \left(1-\frac{2\pi T}{k}\right)^{-1}\right]\right) \label{eq:bhelstat} \\
M_{BH}&=2k e^{2k \phi_{H}}=\frac{q^2k}{16\pi}\left( 1-\frac{2\pi T}{k}\right)^{-1}. \label{eq:bhmass}
\end{align}
Note that the specific heat is positive below the Hagedorn temperature both for the entire spacetime and for the black hole itself.

In deriving the full thermodynamics for this black hole solution we also need the free energy, $F$, of the system. In the classical limit, the free energy is given by
\be
F=-T\ln Z=-T\ln e^{-I}=TI,
\ee
where $I$ is the euclidean action. Since we are going to compare our results with the 0A matrix model, where $q$ is treated as a fixed background parameter \cite{Douglas:2003up}, it is natural to work in the canonical ensemble for the charge $q$. In \cite{Danielsson:2004xf} this was done in detail, with the result
\be \label{eq:totfree}
F_{tot}=\frac{q^2k^2}{8\pi}(\phi_{c}-\phi_{H})+\frac{q^2k}{16\pi}.
\ee
The entropy is now given by
\be \label{eq:bhentropy}
S_{BH}=\frac{1}{T}\left(M_{tot}-F_{tot}\right)=\frac{q^2}{8}\left(1-\frac{2\pi T}{k}\right)^{-1}.
\ee
We can now see that the entropy is proportional to the mass of the black hole, with proportionality constant given by the inverse Hagedorn temperature
\be \label{eq:hagedornrel}
S_{BH}=\beta_{h} M_{BH}.
\ee
It therefore appears as if the low temperature phase is characterized by a stringy density of states. We will explore this further in section \ref{sec:hagedorn}. Naively, (\ref{eq:hagedornrel}) would suggest that the system is at temperature $T_{h}$. But since the \emph{total} mass is $M_{tot}=M_{BH}+M_{el.stat}$, one can verify that
\be
\left.\frac{\partial S_{BH}}{\partial M_{tot}}\right|_{V}=\frac{1}{T},
\ee
where we take the total volume to be $V=\phi_{c}$. We will verify that this is the correct thermodynamic volume later.

In these coordinates it is natural to fix the cutoff at a fixed position in space. From a physical point of view, however, this would mean that the contribution to the energy from the electrostatic potential, decreases as the horizon radius grows larger. Furthermore, as can been seen from eq. (\ref{eq:horiztemp}), the horizon radius goes to infinity in these coordinates as the temperature reaches the Hagedorn temperature $T_{h}=k/2\pi$. At this point, a fixed wall in the $\phi-$coordinate lacks a physical interpretation. 

For these reasons, we will try a physically more sound way of implementing the cutoff. Namely, instead of putting it at a fixed position in space, we will put the cutoff at a fixed position away from the horizon, that is we keep $\phi_{c}-\phi_{H}$ constant. A coordinate system in which this is most naturally done is where the horizon is fixed at the origin. Some features of this coordinate system will be explored in section (\ref{sec:metricsec}).

One immediate consequense of this choice of cutoff is that the free energy (\ref{eq:totfree}) becomes independent of temperature. It is important to note that this is perfectly consistent with a nonvanishing entropy. To make sure this choice is reasonable from a thermodynamic point of view, we will show that it is consistent with the first law of thermodynamics. Note that the charge enters the action (\ref{eq:stringact}) very much like a cosmological constant. This fixed charge ensemble must therefore have a pressure associated with it. Indeed, by standard methods we find
\be
p=-\left.\frac{\partial F_{tot}}{\partial V}\right|_{T}=-\frac{q^2k^2}{8\pi}.
\ee
We are now ready to verify the first law
\be \label{eq:firstlaw}
dF=-SdT-pdV.
\ee
Since the free energy is constant, we have $dF=0$. If we now identify the total volume with the cutoff, i.e. $V\equiv \phi_{c}$, we find that the total volume grows with temperature at the same rate as the horizon does (\ref{eq:horiztemp}). At zero temperature, let $\phi_{c}(T=0)=l_{c}$, then
\be
\phi_{c}=l_{c}-\frac{1}{2k}\ln \left(1-\frac{2\pi T}{k}\right), \label{eq:voltemp}
\ee
so that
\be
dV\equiv d\phi_{c}=\frac{\partial \phi_{c}}{\partial T}dT=\frac{\pi}{k^2}\left(1-\frac{2\pi T}{k}\right)^{-1}dT.
\ee
Given the expressions for the entropy and the pressure, we can now evaluate the right hand side of eq. (\ref{eq:firstlaw})
\begin{align}
-SdT-pdV=&-\frac{q^2}{8}\left(1-\frac{2\pi T}{k}\right)^{-1}dT-\\ &\left(-\frac{q^2k^2}{8\pi}\right)\frac{\pi}{k^2}\left(1-\frac{2\pi T}{k}\right)^{-1}dT =0,
\end{align}
and so the first law is verified. It was important here to identify $\phi_{c}$ as being the correct volume. Another choice might have been $\phi_{c}-\phi_{H}$. However, we only get the correct entropy in the former case, since
\be \label{eq:deriveentropy}
\left.\frac{\partial F_{tot}}{\partial T}\right|_{V}=-S_{BH},
\ee
 only for $V=\phi_{c}$. The first law could also have been verified with a constant, temperature independent, volume. In that case we would have $dV=0$, but then $dF\neq 0$ since there is an explicit temperature dependence through $\phi_{H}$ (\ref{eq:totfree}, \ref{eq:horiztemp}).\footnote{I would like to thank Troels Harmark for discussions on these issues.}

Finally let us comment on the free energy we have (\ref{eq:totfree}). We found that we can make it constant by putting the cutoff at a fixed distance away from the horizon. However, the first of the two terms in (\ref{eq:totfree}) explicitly depends on the cutoff, i.e. the spatial coordinate we choose to use. We know this term is also constant on the matrix model side, and cutoff dependent. The question now is how the cutoffs are related.

It is natural to suspect that the energy cutoff on the matrix model side is related to the total volume on the spacetime side. We know the volume is given by $\phi_{c}$, and that it grows with temperature at the same rate as the horizon radius $\phi_{H}$. From (\ref{eq:horiztemp}) and (\ref{eq:voltemp}) we find
\be \label{eq:constvol}
\phi_{c}-\phi_{H}=-\frac{1}{2k}\ln \left( \frac{q^2}{32\pi}e^{-2kl_{c}} \right).
\ee
We can now replace $\phi_{c}-\phi_{H}$ in (\ref{eq:totfree}) and find
\be \label{eq:totfreel}
F_{tot}=-\frac{q^2k}{16\pi}\ln\left( \frac{q^2}{32\pi}e^{-2kl_{c}} \right)+\frac{q^2k}{16\pi},
\ee
which coincides with the expression for the free energy at zero temperature derived in \cite{Danielsson:2004xf}. Now we see that it is also the correct expression at finite temperature, given our implementation of the cutoff.

As already mentioned, the second term in (\ref{eq:totfreel}), i.e. the one corresponding to the black hole, is {\it analytic} in $q$. This is important to keep in mind when we go to the matrix model, where usually only {\it non-analytic} terms are considered.

\bigskip
\subsection{The black hole metric in a universal form} \label{sec:metricsec}
\bigskip

As mentioned above, a coordinate system in which it is natural to put the cutoff at a fixed position away from the horizon is where the horizon is fixed at the origin. It turns out that the black hole metric in these coordinates simplifies and that some specific features of the system becomes more transparent. So, let us use $(t,r)$, where $r$ is defined through
\be
\phi=\phi_{H}+r.
\ee
The metric component $l(r)$ becomes
\begin{align} \label{eq:metriccompr}
l(r) = 1 - \frac{1}{2k} e^{-2k(\phi_{H}+r)}\left(2k e^{2k \phi_{H}} + \frac{q^2 k^2}{8\pi }r \right).
\end{align}
Simplifying and using eq. (\ref{eq:temp}) we can now replace $\phi_{H}$ in eq. (\ref{eq:metriccompr}) and find
\be \label{eq:metricuniversal}
l(r)=1-e^{-2kr}\left[1+2k\left(1-\frac{2\pi T}{k}\right)r\right],
\ee
where in particular one should note that the dependence on the charge $q$ drops out. This has been noticed both for the near horizon $AdS_{2}$ region of the extremal black hole \cite{Danielsson:2003yi,Thompson:2003fz,Strominger:2003tm}, as well as for the entire extremal black hole itself \cite{Gukov:2003yp}. Here we see that we can express the geometry of the nonextremal black hole in terms of temperature alone, with no reference to the charge. This is surprising since, quite generically, charged black holes are labeled by two parameters; the charge and the deviation away from extremality. In this case, for some reason, we can replace these two parameters by the single parameter $T$.
 Furthermore, we still have a well defined geometry in the limit where $T\rightarrow k/2\pi$. Indeed, in the strict limit we recognize this geometry as Witten's black hole \cite{Witten:1991yr}.
We can put it in a more familiar form by changing coordinates according to
\be
r=\frac{1}{k}\ln\left[\cosh(k\tilde{r})\right].
\ee
In the coordinates $(t,\tilde{r})$ the metric becomes
\be \label{eq:cigar}
ds^{2}=-\tanh^2(k\tilde{r})\text{d}t^2+\text{d}\tilde{r}^2,
\ee
which is precisely the familiar cigar geometry\footnote{Note that this black hole appears here in the limit where the string coupling goes to zero, i.e. Newtons constant is zero \cite{Danielsson:2004xf,Davis:2004xi} and the entropy diverges. Infinite energy is thus required in order to create this black hole.}. It is known that the geometry (\ref{eq:cigar}) admits an exact worldsheet CFT description. Since this geometry appears here as a limit of a more general geometry, one is tempted to speculate that also (\ref{eq:metricuniversal}) admits such a description. 

Let us go back to the coordinates $(t,r)$. We have seen that the metric in these coordinates does not depend on the value of the charge $q$ (as long as $q$ is nonzero). However, the physics still depends on it. Indeed, the charge now sits in the string coupling
\be \label{eq:couplr}
g_{s}^2=e^{2\Phi}=e^{-2k\phi}=e^{-2k(\phi_{H}+r)}=\frac{32\pi}{q^2}\left(1-\frac{2\pi T}{k}\right)e^{-2kr}.
\ee
Interestingly, the charge enters as a \emph{factor}.\footnote{Note that for this reason, curiously, the analogue of the 't Hooft coupling in higher dimensions $\hat{\lambda} \sim g_{s}q$ is in fact also independent of $q$ \cite{Danielsson:2003yi}.} This means that we can rescale the string coupling by $q$ and get a theory which is independent of any parameters, and which in this sense is completely universal. In fact, this may have been anticipated for the following reason. As our starting point we had the action (\ref{eq:stringact}). We can rescale this action according to
\be
\tilde{S}=\frac{1}{q^2}S,
\ee
where then
\be
\tilde{S}= - \int \text{d}^2\text{x} \sqrt{-g} \left[ e^{-2\tilde{\Phi}} \left( \mathcal{R} + 4 (\nabla
   \Phi)^2 + \frac{8}{\al '} \right) -\frac{1}{4\pi \al '} \right].
\ee
The solution to the resulting equations of motion is the metric (\ref{eq:metricuniversal}) as before but now with a coupling
\be
g_{s}^2=e^{2\tilde{\Phi}}=32\pi\left(1-\frac{2\pi T}{k}\right)e^{-2kr}.
\ee
This way of rescaling the action, and get a completely universal solution, is in fact very similar to what was done in \cite{Sen:1995in}. There, the purpose of the rescaling was to find a generic prediction for how the entropy scales with charge, once higher order $\al '$ corrections were included.

This is an important question here as well. As already mentioned, the string coupling can be made arbitrarily small by tuning up $q$, but as can be seen from the metric component (\ref{eq:metricuniversal}), there is no parameter to tune the curvature with. So, quite generically, the curvature close to the horizon will always be of order one\footnote{In this regard, it is interesting to note that, as can easily be verified, the near horizon region of the black hole (\ref{eq:metricuniversal}) interpolates between curvature $\mathcal{R}=-4k^2$ at zero temperature, to curvature $\mathcal{R}=4k^2$ at the Hagedorn temperature. In the intermediate regime where the curvature is small at the horizon, it becomes large at a distance of order one string length away from the horizon.}. Higher order $\al '$-corrections might therefore become important. We can therefore follow the argument in \cite{Sen:1995in} in order to obtain a generic prediction for the scaling of the entropy once higher order $\al '$-corrections are included. Higher curvature corrections to the metric (\ref{eq:metricuniversal}) must be parameter independent and can at most depend on the temperature. In case it still describes a black hole, the entropy at zero temperature can thus only be a numerical constant. Call it $n$. Now, since
\begin{align}
TI &= F = M-TS_{BH}\\ \nonumber
 I&=q^2 \tilde{I},
\end{align}
we find that in the original theory the entropy must be 
\be
S_{BH}=q^2\tilde{S}_{BH} = nq^2.
\ee
This scaling agrees with the prediction of the leading order theory \cite{Davis:2004xb,Danielsson:2004xf,Davis:2004xi}, which furthermore predicts $n=1/8$. Of course, it could be that there is no entropy at tree level in the rescaled theory, once $\al '$-corrections are included. Then clearly the entropy in the original theory also vanishes. It is nevertheless encouraging to see that the dependence on the charge predicted by the leading order theory agrees with this more general argument.

\bigskip
\section{A matrix model black hole?}\label{sec:mmbh}
\bigskip

Let us now turn to the 0A matrix model. In the double scaling limit the theory is reduced to a system of free fermions living in the potential \cite{Douglas:2003up,Maldacena:2005he}
\be
V(x)=-\frac{x^2}{4\al '}+\frac{q^2-\frac{1}{4}}{2x^2}.
\ee
We are interested in calculating the tree level term of the canonical free energy at finite temperature. This has been done many times before, and, as is well known, the leading term is independent of temperature and therefore coincides with the expression for the ground state energy of the system\footnote{Note that this is \emph{not} the case in the 0B theory \cite{Maldacena:2005he} when $q\neq 0$.} (as is the case for the leading order term in spacetime as well (\ref{eq:totfreel})). Thus, for the purpose of finding the leading order term, we only need to calculate the ground state energy, which is given by
\be \label{eq:groundse}
E=\int_{-\epsilon_{c}}^{-\varepsilon}\epsilon\rho(\epsilon)\text{d}\epsilon,
\ee
in which we introduced an energy cutoff $\epsilon_{c}$. The density of states $\rho(\epsilon)$ is given by
\be
\rho(\epsilon)=-\frac{1}{\pi}\text{Im}\sum_{n=0}^{\infty}\frac{1}{E_{n}+\epsilon},
\ee
where $E_{n}$ correspond to single particle energy levels. Since we are mainly interested in the leading order term we can follow \cite{Danielsson:2004xf}, and find
\be
\rho(\epsilon)=-\frac{1}{2\pi k}\ln \left(\frac{q^2}{4}+\frac{\epsilon^2}{k^2}\right)+...,
\ee
which, when plugged into (\ref{eq:groundse}), gives
\be \label{eq:mmtreelevel}
F_{MM}=\frac{q^2k}{16\pi}\left[1-\ln\left(\frac{q^2k^2}{4\epsilon_{c}^2}\right)\right]+...,
\ee
at $\mu=0$. In the last step special care was taken to include analytic terms \cite{Danielsson:2004xf}. In particular (\ref{eq:groundse}) calculates a term
\be
\frac{4\epsilon_{c}^2}{k^2}\ln\left(q^2+\frac{4\epsilon_{c}^2}{k^2}\right),
\ee 
which for large $\epsilon_{c}$ results precisely in the analytic term in (\ref{eq:mmtreelevel}) together with terms independent of $q$ and higher order in $1/\epsilon_{c}$.

We now see that the analytic term found in the matrix model (\ref{eq:mmtreelevel}) coincides precisely with the analytic term on the spacetime side, being due to the black hole itself. The non-analytic term, furthermore, is the same as the one corresponding to the electrostatic energy. On the spacetime side we needed the ADM mass of the spacetime in order to derive the entropy (\ref{eq:bhentropy}). Alternatively, we could find the entropy directly from the constant free energy after identifying the correct thermodynamic volume (\ref{eq:deriveentropy}). Either of these approaches should be possible also on the matrix model side. As for the ADM mass, just as for the free energy, we have seen that it is essential to keep track of the analytic terms in $q$ (\ref{eq:ADMmass}).

While it remains a challenge to find the entropy directly in the matrix model, we can actually use the model to \emph{predict} the first entropy correction of the spacetime system. For this purpose, we write down the non-analytic part of the one-loop term for the free energy (again keeping the cutoff explicit)
\be \label{eq:fmm1loop}
F_{MM}^{1-loop}=\frac{k}{48\pi}\left[1+\left(\frac{2\pi T}{k}\right)^2\right]\ln\left(\frac{q^2k^2}{4\epsilon_{c}^2}\right).
\ee
This term, just like the tree level term, is well known to have a volume dependence (see for example \cite{Klebanov:1991qa}), explaining the presence of the cutoff. In matching the tree level terms corresponding to the electrostatic energy, we can relate the volume cutoffs via the relation
\be \label{eq:volumemap}
\phi_{c}-\phi_{H}=-\frac{1}{2k}\ln \left( \frac{q^2}{32\pi}e^{-2kl_{c}} \right)=-\frac{1}{2k}\ln\left(\frac{q^2k^2}{4\epsilon_{c}^2}\right).
\ee
Comparing this with (\ref{eq:fmm1loop}) immediately suggests that the spacetime one-loop term should read
\be \label{eq:bhfree1loop}
F^{1-loop}=-\frac{k^2}{24\pi}\left[1+\left(\frac{2\pi T}{k}\right)^2\right](\phi_{c}-\phi_{H}),
\ee
from which the quantum corrected entropy can be evaluated according to
\begin{align}
S^{1-loop}=&-\left(\frac{\partial F^{1-loop}}{\partial T}\right)_{\phi_{c}} \nonumber \\
=&\frac{\pi T }{3}(\phi_{c}-\phi_{H}) \label{eq:freegas} \\ 
&-\frac{1}{24}\left[1+\left(\frac{2\pi T}{k}\right)^2\right]\left(1-\frac{2\pi T}{k}\right)^{-1}, \label{eq:bhcorr}
\end{align}
where we used the relation (\ref{eq:voltemp}). The term (\ref{eq:freegas}) has a clear physical interpretation, it is the entropy of a single quantum field (the tachyon) living in the spatial volume $\phi_{c}-\phi_{H}$. More interesting is the second term (\ref{eq:bhcorr}). We interpret it as a quantum corrected contribution to the black hole entropy. It has a temperature dependence very similar to the leading order term (\ref{eq:bhentropy}), but is parametrically smaller by a factor $\sim - q^2$. At zero temperature it is merely a numerical constant $-1/24$.

In relating the cutoffs (\ref{eq:volumemap}), we have on the spacetime side two temperature dependent terms canceling (\ref{eq:constvol}), i.e. $\phi_{c}-\phi_{H}$. But since we are instructed to keep $\phi_{c}$ fixed when deriving the entropy, we effectively have a temperature dependence in the free energy (through $\phi_{H}$), and so we get a nonzero entropy as a result. The relation between the cutoffs might provide a hint to how a similar thing could be done directly in the matrix model. In principle, there should be two canceling terms in the matrix model as well, but since in this case we do not have a clear notion of a horizon radius, it is not clear how to rewrite the expression as we do in the spacetime picture.

Even without a direct matrix model calculation, however, we were able to make a nontrivial prediction for the quantum corrected free energy (\ref{eq:bhfree1loop}), using the matrix model. This term is of an extremely natural form given the expression for the leading order term. It would of course be very interesting to, as a check, calculate the one-loop term directly on the spacetime side.

\bigskip
\section{Speculations on Hagedorn thermodynamics and phase transitions} \label{sec:hagedorn}
\bigskip

\subsection{A phase transition?}
\bigskip

We have seen how the density of states of the black hole grows exponentially with the mass of the black hole. Furthermore, the proportionality constant between the mass and the entropy of the black hole is given by the limiting temperature of the black hole $T_{h}=k/2\pi$. A natural question at this point is what characterizes the physics close to this temperature.

In higher dimensions, with a low temperature density of states like what we found here, typically one expects a phase transition to take place, see for example \cite{Atick:1988si,Aharony:2003sx}. It has been pointed out many times in the literature that in the 2d type 0 theories, this does not happen. The reason is that the theory does not have enough states. Indeed, on the 0A side, the only propagating degree of freedom is the tachyon, while on the 0B side there is only in addition the RR scalar. Being aware of this we will still in this section try to find further support for the idea that there is a phase transition connecting the type 0A and 0B theories. Indeed, the mere existence of a black hole suggests that there is more to the theory than what naively might be expected. For example, the spectrum also contains an infinite number of higher level \emph{discrete} states. Perhaps, just like in critical string theory, these states are responsible for the Hagedorn relation between the mass and the entropy of the black hole.

What in higher dimensions signals the onset of a phase transition is the development of singularities in thermodynamic quantities, as the temperature reaches a critical value. This is precisely what we find here both for the mass (\ref{eq:bhmass}) and the entropy (\ref{eq:bhentropy}) of the black hole. Even though we have a constant free energy (\ref{eq:totfree}), since we keep $\phi_{c}-\phi_{H}$ fixed, it has an effective logarithmic divergence in operations where the volume $\phi_{c}$ is kept fixed (\ref{eq:deriveentropy}).

In systems where phase transitions happen, these types of singularities can be understood in essentially two ways. One is by the exponential mass dependence for the density of states $\rho(M)\sim e^{\beta_{h}M}$. If we formally write the logarithm of the thermal partition function as
\be
\ln (Z(T))\sim \int dM \rho(M) \mbox{ }e^{-\beta M}\sim \int dM e^{\beta_{h}M-\beta M},
\ee 
we see that it diverges for temperatures larger than the Hagedorn temperature $T_{h}$. Since we found here that the density of states grows exponentially with the mass of the black hole, we can explain the divergences we have found precisely in this way. Note however that the degeneracy does not grow with the total ADM mass of the spacetime, but only with the mass of the black hole. In practice what one does is to first integrate out the vacuum part, which in this case corresponds to the electrostatic energy, and then separately take care of the thermal part, here the black hole.

As mentioned, the divergence can also be explained from an alternative point of view. Namely, if in the spectrum a mode becomes tachyonic at some critical temperature, the partition function will diverge there due to the appearance of the negative mass squared term. This tachyonic mode can also be considered as the dynamical mechanism that takes the system from one phase to another. 

As is well known (see for example \cite{Atick:1988si}), this happens in critical string theory for strings winding the thermal circle, when the circle is small enough. In our case, however, the lowest lying state, the so called tachyon, is \emph{not} tachyonic. And winding the thermal circle will only make it heavier. However, it has been shown that the mass of the originally massless tachyon depends in a  nontrivial way on the black hole background. In case of the \emph{extremal} black hole, the mass is lifted to positive mass \cite{Thompson:2003fz,Strominger:2003tm}. But let us repeat the analysis of \cite{Thompson:2003fz,Strominger:2003tm} in case of the \emph{nonextremal} black hole. Quite remarkably, we will find that precisely at the Hagedorn temperature, the tachyon becomes truly tachyonic.

In order to obtain the linearized equation of motion for the tachyon $T$, we need to generalize the action (\ref{eq:stringact}) to leading order in $T$. With only one one RR flux turned on, it is given by \cite{Douglas:2003up}
\begin{align}
S = - \int \text{d}^2 x \sqrt{-g} \left[  e^{-2\Phi} \left(   \frac{8}{\al '}+ \mathcal{R} + 4 (\nabla
   \Phi)^2 -\frac{1}{2}(\nabla T)^{2} + \frac{1}{\al '}T^{2} \right)\right. \\
 -(1+2T+2T^{2}+...)\left.\frac{q^2}{4\pi \al '} \right]
\end{align}
The equation of motion for the rescaled tachyon $e^{-\Phi}T$ becomes
\be
\left[ \nabla^{2}+\nabla^{2}\Phi-(\nabla \Phi)^{2}+\frac{2}{\al '}-\frac{q^{2}}{\pi \al '}e^{2\Phi}\right](e^{-\Phi}T)=\frac{q^{2}}{2\pi \al '}e^{\Phi},
\ee
which has the form of a sourced\footnote{The source can be made to go away by including the second flux $\tilde{q}$ with equal size as $q$ \cite{Thompson:2003fz,Strominger:2003tm}, i.e. $\tilde{q}=q$. But then one must also worry about the additional $q\tilde{q}$ strings which should be added to the spacetime \cite{Maldacena:2005he}.} Klein-Gordon equation with a mass-term
\be
m_{T}^{2}=\nabla^{2}\Phi-(\nabla \Phi)^{2}+\frac{2}{\al '}-\frac{q^{2}}{\pi \al '}e^{2\Phi}.
\ee
Using (\ref{eq:phieqmot}) the mass term can be written as
\be
m_{T}^{2}=\frac{\mathcal{R}}{4}+\frac{q^2}{\pi \al '}e^{2\Phi}.
\ee
In the coordinate system $(r,t)$ $e^{2\Phi}$ is given by (\ref{eq:couplr}) and the Ricci scalar is
\be
\mathcal{R}=\frac{8}{\al '}e^{-2kr}\left[2kr-\frac{2r}{R}+ \frac{2}{kR} -1\right],
\ee
where $R$ is the radius of the thermal circle, i.e. $\beta=2\pi R$. We can now express the mass squared of the tachyon in terms of position $r$ and the radius of the thermal circle according to
\be \label{eq:massT}
m_{T}^{2}=\frac{2}{\al '}e^{-2kr}\left[ 2kr -\frac{2r}{R}-\frac{14}{kR}+ 15 \right].
\ee
This is a generalization of the mass formula for the tachyon found in \cite{Thompson:2003fz}, which considered the near horizon region at zero temperature. As a check, it is easy to verify that (\ref{eq:massT}) reduces to the result in \cite{Thompson:2003fz} in the limit $r\rightarrow 0$ and $R\rightarrow \infty$.

Now, note that (\ref{eq:massT}) is always positive outside the horizon (i.e. when $r>0$) as long as the temperature is below the Hagedorn temperature, corresponding to $R>1/k$. However, as soon as $R<1/k$ there is \emph{always} a point outside the horizon beyond which the tachyon becomes truly tachyonic all the way out to infinity. As promised, we have found a tachyonic mode which appears \emph{precisely} at the Hagedorn temperature, thus further supporting the idea that the system goes through a phase transition at that point. We find this result quite suggestive.\footnote{We should point out that the tachyon considered here is of a rather different nature than the more familiar winding tachyons. This tachyon more resembles spacetime dependent ``quasilocalized'' tachyons considered in for example \cite{Horowitz:2006mr}. In this regard, note that \emph{inside} the horizon ($r<0$), \emph{below} the Hagedorn temperature, beyond a certain point the mass squared (\ref{eq:massT}) becomes \emph{negative} with an exponentially growing value.}

\bigskip
\subsection{The high temperature phase}
\bigskip

We continue here by making a preliminary investigation of the physics above the Hagedorn temperature. Since this temperature corresponds to the self-dual radius of the T-dual 0A and 0B theories \cite{Danielsson:2004ti,Maldacena:2005he,Danielsson:2005uz}, the physics above the Hagedorn temperature in 0A, is thus more properly described at temperatures below the Hagedorn temperature in 0B. In this sense the 0A theory is the low temperature phase, while the high temperature phase is 0B. We perform this analysis on the matrix model side. We will be brief here and refer to \cite{Maldacena:2005he} for further details.

In the 0B theory the potential is given by
\be
V(x)=-\frac{x^2}{4\al '},
\ee
and is filled by fermions on both sides. Filling the two sides to different levels is related to the instanton charge of the theory \cite{Douglas:2003up}. In the Euclidean formulation the difference in levels become imaginary shifts of the Fermi surface \cite{Maldacena:2005he}. This means that in the grand canonical ensemble the theory has two Fermi levels given by $\mu_{B}+iq/R_{B}$ and $\mu_{B}-iq/R_{B}$ respectively. Let us compare the tree level and one-loop expressions of the free energies in this ensemble for the type 0A and 0B theories at $\mu_{A}=\mu_{B}=0$. We will ignore the analytic terms here and only focus on the ground state energy and the degrees of freedom in the gas at temperature $T=1/(2\pi R)$.

In general in $2d$, one has 
\be \label{eq:measure}
\frac{\ln {Z}}{V}=a\frac{R}{\al '}+\frac{b}{R},
\ee 
where the dimensionless constant $a$ is a measure of the ground state energy, and $b$ is a measure of the degrees of freedom of the system. Usually both these terms come from one-loop calculations. But as we have seen, the ground state of the 0A theory receives contributions already at tree level. From the results of section (\ref{sec:mmbh}) we can write the (grand) canonical free energy at $\mu_{A}=0$ as\footnote{At $\mu_{A}=0$ the expressions for the canonical and grand canonical free energies are obviously the same. This is not so in the 0B theory since at $\mu_{B}=0$ there are still the nonzero imaginary shifts in the Fermi levels. We denote the grand canonical free energies by a tilde.}
\be
-\frac{2\pi R_{A}\tilde{F}_{A}}{V}=\frac{\ln (Z_{A})}{V}=-\frac{1}{2}\left(q^2-\frac{1}{3}\right)\frac{R_{A}}{\al '}+\frac{1}{12R_{A}}.
\ee  
From this we can immediately find the corresponding 0B expression by using the T-dual operation at $\mu=0$ \cite{Maldacena:2005he}
\be
R_{A}R_{B}=\al '.
\ee 
Alternatively, we can read it off directly from \cite{Maldacena:2005he}. The result is the same
\be
-\frac{2\pi R_{B}\tilde{F}_{B}}{V}=\frac{\ln (Z_{B})}{V}=\frac{R_{B}}{12\al '}-\frac{1}{2}\left(q^2-\frac{1}{3}\right)\frac{1}{R_{B}}.
\ee
Before interpreting this result we must Legendre transform the expression for the grand canonical free energy to the canonical free energy. Essentially, we have two sides with Fermi levels $iq/R_{B}$ and $-iq/R_{B}$ respectively. Legendre transforming these two sides results in a \emph{sign flip} of the $q^2$-term.

The main point now is the following: After transforming to the canonical ensemble, the term measuring the number of degrees of freedom (the constant $b$ in (\ref{eq:measure})) makes a sudden jump as one goes from 0A to 0B
\be
b:\quad \frac{1}{12} \rightarrow \frac{1}{2}\left(q^2+\frac{1}{3}\right).
\ee
We know that in 0A the $1/12$ corresponds to a single scalar degree of freedom, i.e. the tachyon. We therefore see that the precise number of real scalar degrees of freedom $n_{dof}\equiv 12b$ makes the following jump at the Hagedorn temperature
\be
n_{dof}: \quad 1 \rightarrow 6q^2+2.
\ee
The 2 is easy to understand; it accounts for the tachyon and the RR scalar \cite{Douglas:2003up,Takayanagi:2003sm}. But now we find in addition a tree level contribution $6q^2$. In 0A it was the ground state energy which received a tree level contribution, in 0B it is instead the number of degrees of freedom. Even though, by T-duality, this should be the case, it is quite mysterious and we do not have a clear understanding of this result. Should one interpret the $6q^2$ as true physical real scalar degrees of freedom of the system, and if so, what are they?\footnote{We note that there are additional massless solitonic states in 0B \cite{DeWolfe:2003qf}, which do not exist in 0A. From the point of view of the 0A theory, these states would appear as new UV degrees of freedom.}

Surprising as it may be, this is strikingly similar to the \emph{deconfined} phase of large $N$ gauge theories. Typically, in confining $U(N)$ gauge theories the low temperature phase is characterized by a Hagedorn density of states \cite{Aharony:2003sx} and a gas of $\mathcal{O}(1)$ degrees of freedom, corresponding to gauge singlets. The ground state energy is of order $N^2$. In the high temperature deconfined phase, the gas has of order $N^2$ degrees of freedom, since now all the individual gluons can contribute. This clearly resembles the jump we have seen in the degrees of freedom as one enters the high temperature 0B phase.

Another interesting analogue to confining theories is the \emph{linear} Coulomb potential we have in 0A. It looks very much like a flux tube. A defining property of confining theories is that it takes infinite energy to add a single external quark to the theory. The reason is that the flux tube (or ``QCD-string'') goes from the quark all the way to infinity since it has nowhere else to end. In 0A we see that it takes infinite energy to add a single D0 brane, since in this case the electric potential grows linearly all the way to infinity. Perhaps one can think of this ``tube'' as condensing at the Hagedorn temperature, and resulting in what we found on the 0B side?\footnote{The possibility of a phase transition in the type 0 theories, similar to that in large $N$ gauge theories, have also been suggested in \cite{Aharony:2003sx}.}

Let us address one more point before concluding. In other examples of the gauge/gravity correspondence it is the high temperature deconfined phase which corresponds to the black hole \cite{Witten:1998qj,Witten:1998zw}. Here we found the black hole in the low temperature 0A regime. There is no contradiction here. In AdS, the specific heat for black holes is only positive for so called large black holes \cite{Hawking:1982dh}. These black holes can only exist in stable equilibrium with the heat bath if the heat bath temperature is large enough. For the 0A black hole, on the other hand, already the low temperature Hagedorn phase has a positive specific heat, thus describing a stable black hole system below the Hagedorn temperature.

\bigskip
\section{Concluding remarks}\label{sec:conclude}
\bigskip

We have found that the free energy calculated on the matrix model side agrees with the one calculated on the spacetime side, also at finite temperature. The analytic term corresponding to the black hole itself as well as the term representing the electrostatic potential, are both exactly reproduced by the matrix model.

Furthermore, using our prescription for how the volume outside the horizon is related to matrix model expressions, we could use the one-loop term of the matrix model free energy to predict the next to leading order expression for the entropy of the black hole spacetime. The result revealed two distinct contributions. One due to a bosonic gas between the horizon and the cutoff. The other due to the black hole itself with a temperature dependence similar to the leading order term, but parametrically smaller by a factor $-q^2$.

Our findings are clearly compatible with the existence of a matrix model black hole. We have not, however, been able to derive the entropy directly in the matrix model. This is still an important open problem. We have emphasized the importance of keeping track of the analytic terms in the analysis. This is in general difficult when cutoff dependent and non-analytic terms are present. For the free energy we could do this, and it should also be possible in the calculation of the total mass at finite temperature. This is the analogue of the ADM mass calculation on the spacetime side.

Results both from the spacetime analysis and the matrix model suggest that the $2d$ type 0 theory should be viewed as one theory with two phases, separated by a phase transition; 0A being the low temperature phase, and, by T-duality, 0B the high temperature phase. The low temperature phase describes a black hole and is characterized at finite temperature by a Hagedorn density of states and a gas with one single degree of freedom, i.e. the tachyon. The fact that the black hole exists in the low temperature phase is simply a consequence of the positive specific heat for the black hole in this regime. This is in contrast to for example AdS black holes, which do not exist at low temperature.

The high temperature 0B phase more resembles a ``deconfined'' phase where now the gas, apart from the tachyon and the RR scalar, apparently also consists of an additional $6q^2$ degrees of freedom. As explained in the main text, by T-duality this result should be expected, given the 0A expression. On the other hand we only expect the mentioned two degrees of freedom in the 0B theory. It would clearly be very interesting to interpret this result on the spacetime side.

As further evidence for a phase transition at the Hagedorn temperature, we have identified the tachyonic mode which appears at that point. The mass of the so called tachyon in the 0A phase depends in a nontrivial way on the background, and we found that it actually becomes tachyonic precisely at the Hagedorn temperature (\ref{eq:massT}).

Clearly, it is too soon to draw any definite conclusions as there are many important issues yet to be settled. Most pressing, perhaps, is the lack of a direct calculation of the black hole entropy in the matrix model. But we have also found some new interesting results. This makes us feel there is some exciting physics in these models, yet to be fully appreciated.

\bigskip
\begin{center}
\section*{Acknowledgements}
\end{center}
\bigskip
I would like to thank Troels Harmark, Shinji Hirano and Marcel Vonk for discussions. I would also like to thank Ulf Danielsson for discussions and in particular for pointing out the importance of analytic terms for me.


\bigskip
\begin{thebibliography}{99}

\bibitem{Douglas:2003up}
  M.~R.~Douglas, I.~R.~Klebanov, D.~Kutasov, J.~Maldacena, E.~Martinec and N.~Seiberg,
  ``A new hat for the c = 1 matrix model,''
  arXiv:hep-th/0307195.

\bibitem{Takayanagi:2003sm}
  T.~Takayanagi and N.~Toumbas,
  ``A matrix model dual of type 0B string theory in two dimensions,''
  JHEP {\bf 0307}, 064 (2003)
  [arXiv:hep-th/0307083].

\bibitem{Berkovits:2001tg}
  N.~Berkovits, S.~Gukov and B.~C.~Vallilo,
  ``Superstrings in 2D backgrounds with R-R flux and new extremal black
  holes,''
  Nucl.\ Phys.\ B {\bf 614}, 195 (2001)
  [arXiv:hep-th/0107140].

\bibitem{Gukov:2003yp}
  S.~Gukov, T.~Takayanagi and N.~Toumbas,
  ``Flux backgrounds in 2D string theory,''
  JHEP {\bf 0403}, 017 (2004)
  [arXiv:hep-th/0312208].

\bibitem{Davis:2004xb}
  J.~L.~Davis, L.~A.~Pando Zayas and D.~Vaman,
  ``On black hole thermodynamics of 2-D type 0A,''
  JHEP {\bf 0403}, 007 (2004)
  [arXiv:hep-th/0402152].

\bibitem{Danielsson:2004xf}
  U.~H.~Danielsson, J.~P.~Gregory, M.~E.~Olsson, P.~Rajan and M.~Vonk,
  ``Type 0A 2D black hole thermodynamics and the deformed matrix model,''
  JHEP {\bf 0404}, 065 (2004)
  [arXiv:hep-th/0402192].

\bibitem{Davis:2004xi}
  J.~L.~Davis and R.~McNees,
  ``Boundary counterterms and the thermodynamics of 2-D black holes,''
  JHEP {\bf 0509}, 072 (2005)
  [arXiv:hep-th/0411121].

\bibitem{Danielsson:2003yi}
  U.~H.~Danielsson,
  ``A matrix model black hole: Act II,''
  JHEP {\bf 0402}, 067 (2004)
  [arXiv:hep-th/0312203].

\bibitem{Thompson:2003fz}
  D.~M.~Thompson,
  ``AdS solutions of 2D type 0A,''
  Phys.\ Rev.\ D {\bf 70}, 106001 (2004)
  [arXiv:hep-th/0312156].

\bibitem{Strominger:2003tm}
  A.~Strominger,
  ``A matrix model for AdS(2),''
  JHEP {\bf 0403}, 066 (2004)
  [arXiv:hep-th/0312194].

\bibitem{Ho:2004qp}
  P.~M.~Ho,
  ``Isometry of AdS(2) and the c = 1 matrix model,''
  JHEP {\bf 0405}, 008 (2004)
  [arXiv:hep-th/0401167].

\bibitem{Aharony:2005hm}
  O.~Aharony and A.~Patir,
  ``The conformal limit of the 0A matrix model and string theory on AdS(2),''
  arXiv:hep-th/0509221.

\bibitem{Jevicki:1993zg}
  A.~Jevicki and T.~Yoneya,
  ``A Deformed matrix model and the black hole background in two-dimensional
  string theory,''
  Nucl.\ Phys.\ B {\bf 411}, 64 (1994)
  [arXiv:hep-th/9305109].

\bibitem{Danielsson:1993wq}
  U.~H.~Danielsson,
  ``A Matrix model black hole,''
  Nucl.\ Phys.\ B {\bf 410}, 395 (1993)
  [arXiv:hep-th/9306063].

\bibitem{Demeterfi:1993sj}
  K.~Demeterfi and J.~P.~Rodrigues,
  ``States and quantum effects in the collective field theory of a deformed
  matrix model,''
  Nucl.\ Phys.\ B {\bf 415}, 3 (1994)
  [arXiv:hep-th/9306141].

\bibitem{Maldacena:2005he}
  J.~Maldacena and N.~Seiberg,
  ``Flux-vacua in two dimensional string theory,''
  JHEP {\bf 0509}, 077 (2005)
  [arXiv:hep-th/0506141].

\bibitem{Friess:2004tq}
  J.~J.~Friess and H.~L.~Verlinde,
  ``Hawking effect in 2-D string theory,''
  arXiv:hep-th/0411100.

\bibitem{Karczmarek:2004bw}
  J.~L.~Karczmarek, J.~Maldacena and A.~Strominger,
  ``Black hole non-formation in the matrix model,''
  arXiv:hep-th/0411174.

\bibitem{Martinec:2004qt}
  E.~Martinec and K.~Okuyama,
  ``Scattered results in 2D string theory,''
  JHEP {\bf 0410}, 065 (2004)
  [arXiv:hep-th/0407136].

\bibitem{Sen:2004yv}
  A.~Sen,
  ``Symmetries, conserved charges and (black) holes in two dimensional  string
  theory,''
  JHEP {\bf 0412}, 053 (2004)
  [arXiv:hep-th/0408064].

\bibitem{Gross:1990md}
  D.~J.~Gross and I.~R.~Klebanov,
  ``Vortices And The Nonsinglet Sector Of The C = 1 Matrix Model,''
  Nucl.\ Phys.\ B {\bf 354}, 459 (1991).

\bibitem{Witten:1991yr}
  E.~Witten,
  ``On string theory and black holes,''
  Phys.\ Rev.\ D {\bf 44}, 314 (1991).

\bibitem{Sen:1995in}
  A.~Sen,
  ``Extremal black holes and elementary string states,''
  Mod.\ Phys.\ Lett.\ A {\bf 10}, 2081 (1995)
  [arXiv:hep-th/9504147].

\bibitem{Kazakov:2000pm}
  V.~Kazakov, I.~K.~Kostov and D.~Kutasov,
  ``A matrix model for the two-dimensional black hole,''
  Nucl.\ Phys.\ B {\bf 622}, 141 (2002)
  [arXiv:hep-th/0101011].

\bibitem{Klebanov:1991qa}
  I.~R.~Klebanov,
  ``String theory in two-dimensions,''
  arXiv:hep-th/9108019.

\bibitem{Atick:1988si}
  J.~J.~Atick and E.~Witten,
  ``The Hagedorn Transition And The Number Of Degrees Of Freedom Of String
  Theory,''
  Nucl.\ Phys.\ B {\bf 310}, 291 (1988).

\bibitem{Aharony:2003sx}
  O.~Aharony, J.~Marsano, S.~Minwalla, K.~Papadodimas and M.~Van Raamsdonk,
  ``The Hagedorn / deconfinement phase transition in weakly coupled large N
  gauge theories,''
  Adv.\ Theor.\ Math.\ Phys.\  {\bf 8}, 603 (2004)
  [arXiv:hep-th/0310285].

\bibitem{Horowitz:2006mr}
  G.~T.~Horowitz and E.~Silverstein,
  ``The inside story: Quasilocal tachyons and black holes,''
  arXiv:hep-th/0601032.

\bibitem{Danielsson:2004ti}
  U.~H.~Danielsson, M.~E.~Olsson and M.~Vonk,
  ``Matrix models, 4D black holes and topological strings on non-compact
  Calabi-Yau manifolds,''
  JHEP {\bf 0411}, 007 (2004)
  [arXiv:hep-th/0410141].

\bibitem{Danielsson:2005uz}
  U.~H.~Danielsson, N.~Johansson, M.~Larfors, M.~E.~Olsson and M.~Vonk,
  ``4D black holes and holomorphic factorization of the 0A matrix model,''
  arXiv:hep-th/0506219.

\bibitem{DeWolfe:2003qf}
  O.~DeWolfe, R.~Roiban, M.~Spradlin, A.~Volovich and J.~Walcher,
  ``On the S-matrix of type 0 string theory,''
  JHEP {\bf 0311}, 012 (2003)
  [arXiv:hep-th/0309148].

\bibitem{Witten:1998qj}
  E.~Witten,
  ``Anti-de Sitter space and holography,''
  Adv.\ Theor.\ Math.\ Phys.\  {\bf 2}, 253 (1998)
  [arXiv:hep-th/9802150].

\bibitem{Witten:1998zw}
  E.~Witten,
  ``Anti-de Sitter space, thermal phase transition, and confinement in  gauge
  theories,''
  Adv.\ Theor.\ Math.\ Phys.\  {\bf 2}, 505 (1998)
  [arXiv:hep-th/9803131].

\bibitem{Hawking:1982dh}
  S.~W.~Hawking and D.~N.~Page,
  ``Thermodynamics Of Black Holes In Anti-De Sitter Space,''
  Commun.\ Math.\ Phys.\  {\bf 87}, 577 (1983).

\end{thebibliography}
\end{document}